\documentclass[a4paper, oneside, twocolumn, notitlepage, 10pt]{extarticle_ecoc}
\usepackage{ecoc}
\usepackage{amsmath,amssymb}
\usepackage{readarray}
\usepackage{pgfplots} 
\usepackage{adjustbox}
\usepackage{pgffor}
\pgfplotsset{compat=1.16}
\usepgfplotslibrary{colorbrewer}
\usepgfplotslibrary{fillbetween}

\begin{document}
\selectlanguage{english}    % Standard Language

%-------------------------------------------------- Title -----------------------------------------------------%

\title{2.4-THz Bandwidth Optical Coherent Receiver Based on a Photonic Crystal Microcomb}%

%------------------------------------------------- Authors-----------------------------------------------------%

\author{
    Callum Deakin\textsuperscript{(1)}, Jizhao Zang\textsuperscript{(2,3)},
    Xi Chen\textsuperscript{(1)}, Di Che\textsuperscript{(1)}, Lauren Dallachiesa\textsuperscript{(1)}, \\ Brian Stern\textsuperscript{(1)}, Nicolas K. Fontaine\textsuperscript{(1)}, Scott Papp\textsuperscript{(2,3)}
}
\maketitle                  % Create title and author

%------------------------------------------ Description of Authors ----------------------------------------------%

\begin{strip}
    \begin{author_descr}

        \textsuperscript{(1)} Nokia Bell Labs, 600 Mountain Ave, Murray Hill, NJ, USA
        \textcolor{blue}{\uline{callum.deakin@nokia-bell-labs.com}}

        \textsuperscript{(2)} Time and Frequency Division, National Institute of Standards and Technology, Boulder, Colorado, USA

        \textsuperscript{(3)} Department of Physics, University of Colorado, Boulder, Colorado, USA

    \end{author_descr}
\end{strip}

% \setstretch{1.1}
%-------------------------------------------------- Footnote -------------------------------------------------------%
\renewcommand\footnotemark{}
\renewcommand\footnoterule{}
%\let\thefootnote\relax\footnotetext{text}

%-------------------------------------------------- Abstract ---------------------------------------------------------%

\begin{strip}
    \begin{ecoc_abstract}
        We demonstrate a spectrally-sliced single-polarization optical coherent receiver with a record 2.4-THz bandwidth, using a 200-GHz tantalum pentoxide photonic crystal microring resonator as the local oscillator frequency comb. \textcopyright2024 The Author(s)
    \end{ecoc_abstract}
\end{strip}

%-------------------------------------------------- Introduction Section -------------------------------------------------------%

\section{Introduction}

It is anticipated that the increasing traffic in optical networks will precipitate a shift away from the current wavelength-routing-only paradigm~\cite{winzer2017scaling,schmogrow2022solving}. In this scenario, scaling is achieved through both wavelength and spatial parallelism, and THz bandwidth optical superchannels are generated, routed, and detected as an single entity. This will require optical transceivers with several THz of bandwidth (or higher) while maintaining a small footprint, low cost, and low power consumption. However, current optical transceivers are limited in bandwidth to around 100 GHz, over an order of magnitude below what is required to fill the optical C-band (4.4 THz), and several orders of magnitude below the exploitable bandwidth of the optical fiber channel ($>$37.6~THz~\cite{PuttnamOESCLU}).  The main culprits of this bandwidth limitation are the DACs and ADCs, which are both fundamentally limited by clock jitter and the switching speed of the integrated circuit technology~\cite{walden1999analog}. On the other hand, scaling through simple optical parallelism (i.e., a laser bank) makes it challenging to maintain a ultra compact footprint and a cost efficient solution, and results in a spectral efficiency penalty compared to fully synchronised manipulation of the optical waveform~\cite{fontaine2013fiber,temprana2015overcoming,sohanpal2023impact}.

Optical frequency combs can offer solutions to these issues. Combs based on microring resonators can achieve exceptionally low timing jitter in a chip-scale footprint and can be used to build optical coherent receivers that stitch together multiple electronic sub-receivers to achieve a higher net bandwidth, bypassing the speed limitations of the electronic integrated circuits to achieve demonstrated bandwidths of up to 610 GHz~\cite{fontaine2010real,fontaine2013fiber,shi2017246,drayss2023non}, and can be fully integrated in silicon photonics~\cite{drayss2024non}. However, conventional microcombs typically operate with pump conversion efficiencies of less than 5\% and generate significant spectral content outside of the frequencies of interest~\cite{bao2014nonlinear}, which is incompatible with transceivers' strict power dissipation requirements. Furthermore, fast frequency or power sweeps of the pump laser are required to access the soliton state due to intra-cavity thermo-optic effects, which prevent the generation of a frequency comb unless the transition to soliton state can occur much faster than the material thermal lifetime~\cite{li2017stably}. This requires fast (MHz) control of the pump laser which adds significant design effort and complexity to the transceivers.

In this paper, we build a record high bandwidth 2.4-THz optical coherent receiver with a dark soliton microcomb generated by a dispersion-engineered photonic crystal resonator (PhCR) with a pump conversion efficiency of $>$20\%. Unlike conventional resonator designs, the soliton state can be initiated with a simple slow (sub-Hz speed) pump frequency sweep and near unit ($>$86\%) pump conversion efficiencies~\cite{zang2022near,zang2024laserpower}.

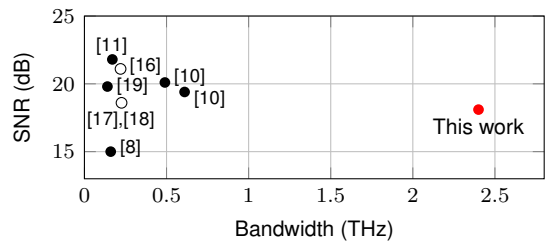
\begin{figure}[hb]
%\vspace{-0.5cm}
   \centering
    \begin{tikzpicture}[trim axis left,trim axis right]

\begin{axis} [ylabel= SNR (dB), 
              xlabel=Bandwidth (THz),
              xmin=0,xmax=2.8,
              ymax=25,
              ymin=13,
              height=0.49\linewidth,
              grid=both,
              width=\linewidth,
              cycle list/Dark2,
              ylabel near ticks,
              legend pos=south west,
              legend columns =2,
              clip mode=individual,
              log ticks with fixed point,
              legend style={font=\footnotesize},
              label style={font=\footnotesize},
              tick label style={font=\footnotesize},]

%\node at (axis cs:300, 45) {(a)};

\fill (0.490,20.1) circle[radius=2pt] node[anchor=west,yshift=1]{{\textcolor{black}{\cite{drayss2023non}}}};;
\fill (0.610,19.4) circle[radius=2pt] node[anchor=west,yshift=-4]{{\textcolor{black}{\cite{drayss2023non}}}};;
\fill (0.160,15) circle[radius=2pt] node[anchor=west]{{\textcolor{black}{\cite{fontaine2010real}}}};;
\draw[black] (0.220,21.1) circle[radius=2pt] node[anchor=west]{{\textcolor{black}{\cite{geiger2023performance}}}};;
\fill[red] (2.4,18.1) circle[radius=2pt] node[anchor=north]{{\textcolor{black}{\footnotesize{This work}}}};;
\draw[black] (0.226,18.6) circle[radius=2pt] node[anchor=north]{{\textcolor{black}{\cite{che2023,yamazaki}}}};;
\fill (0.140,19.8) circle[radius=2pt] node[anchor=west,yshift=-1]{{\textcolor{black}{\cite{fang2022optical}}}};;
\fill (0.170,21.8) circle[radius=2pt] node[anchor=south,yshift=-4]{{\textcolor{black}{\cite{drayss2024non}}}};;

%\addplot [mark=o,only marks,color=black]  coordinates {( -0.50790625,16.7) };
%\addlegendentry{Multi-carrier CoRx};

\end{axis}
\end{tikzpicture}
    \caption{High bandwidth optical coherent receiver demonstrations, including single carrier (open circles) and comb-based optical multiplexing schemes (solid circles).}
    \label{fig:comparison}
 %   \vspace{-0.5cm}

\end{figure}

\section{Photonic crystal microring resonator}
\begin{figure*}[tb]
%\vspace{-0.5cm}
   \centering
    \includegraphics[width=\linewidth]{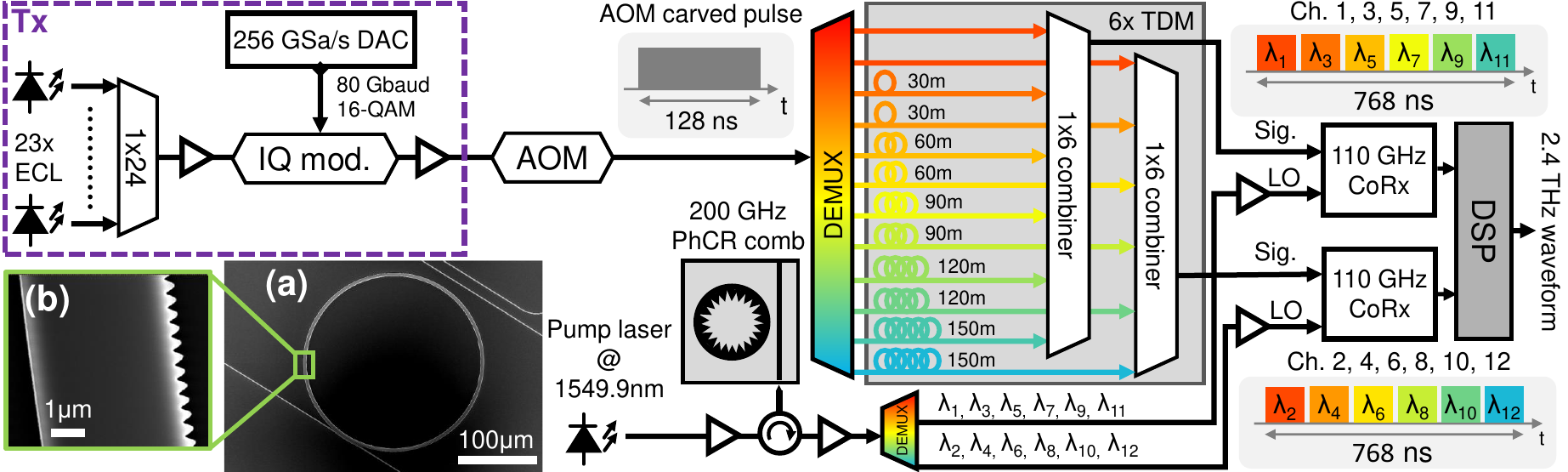}
    \caption{Experimental setup. LO, local oscillator. Insets: scanning electron microscope (SEM) images of (a) the fabricated ring resonator and (b) detail of the ring width oscillations on the inner wall of the ring resonator.}
    \label{experiment_setup}
 %   \vspace{-0.5cm}

\end{figure*}
The PhCR, shown in Fig.~\ref{experiment_setup}(a)/(b), is fabricated in tantalum pentoxide, which offers lower residual stress, higher nonlinear index, and smaller thermo-optic coefficient compared to silicon nitride~\cite{jung2021tantala}.  The PhCR has normal group velocity dispersion as the dark solitons that form in this regime offer better spectral flatness and higher conversion efficiency. To control the PhCR dispersion, we design uniform ring width oscillations (see Fig.\ref{experiment_setup}(b)) to induce a photonic bandgap that manifests as mode splitting at the desired resonance~\cite{lucas2023tailoring}. The lower-frequency resonance of the split mode meets the phase-matching condition, allowing for spontaneous soliton formation~\cite{Su-peng2021}. Solitons can then be initiated by pumping the lower-frequency resonance of the split mode, followed by a slow (sub-Hz)  frequency detuning sweep of a few GHz to generate a 200-GHz spaced frequency comb centered at 193.47~THz. Unlike conventional microring combs, the generation of a soliton in our PhCR is very insensitive to the speed of the sweep because the soliton is thermally stable in the resonator, and it can be dispersion-engineered such that comb modes only appear in the desired spectral range.  Light is edge-coupled in and out of the ring resonator chip via lensed fibers with coupling losses of approximately 4.5~dB/facet, which can be reduced to around 1.5~dB/facet if an oxide cladding is applied. The on-chip pump power threshold for soliton generation was measured to be 18.2~dBm, therefore the pump laser is amplified to 23~dBm to compensate for the coupling losses. 

\section{Experimental setup}

The full experimental setup used to achieve a 2.4~THz detection bandwidth is shown in Fig.~\ref{experiment_setup}. A circulator is used to obtain the counter propagating soliton frequency comb and the observed frequency comb spectrum is shown in Fig.~\ref{fig:results}(a). The 12 comb lines indicated by the shaded area in Fig.~\ref{fig:results}(a) are used as the local oscillators for our spectrally sliced coherent receiver. The power per line varies from -0.5 dBm to -12.7 dBm after coupling to the fiber: the aggregate power of the 12 lines is 6.7~dBm ($\approx$11.2~dBm on chip). At the receiver, a wavelength selective switch (WSS) is used to select and equalize the 12 comb lines. In a practical implementation, an on-chip demux device (e.g. an arrayed waveguide grating) should be used with the PhCR engineered to an even flatter power profile and equalised by on-chip attenuations instead of a WSS. Ideally, the demuxed 12 lines would be sent to 12 individual 100-GHz receivers for coherent detection. Since we do not have 12 100-GHz receivers, we emulated the required receivers through a time division multiplexing (TDM) receiver technique~\cite{van2014time}. The comb lines are split into two groups, the odd ($\lambda_1, \lambda_3,\dots, \lambda_{11}$) and even channels ($\lambda_2, \lambda_4,\dots, \lambda_{12}$), which are then fed to the local oscillator ports of two 110 GHz coherent receivers with approximately 10~dBm per comb line after amplification by an erbium doped fiber amplifier (EDFA). 

Meanwhile, a 2.4-THz test signal is generated by modulating 23 external cavity lasers (ECLs) with 80-GBaud single-polarization 16-QAM signals via an IQ modulator. The ECLs are spaced at about 100 GHz and positioned such that a modulated signal is covering every channel overlap and are spread over a 2.4~THz bandwidth. To enable the TDM technique, the signal is gated into a 128-ns pulse via a 200~MHz acoustic optic modulator (AOM), which is then split into the 12$\times$200~GHz bandwidth slices by a WSS. The 12 channels are divided into 2 groups, the odd (centered at comb lines $\lambda_1, \lambda_3, \dots, \lambda_{11}$) and even (centered at comb lines $\lambda_2, \lambda_4, \dots, \lambda_{12}$) channels, to be sent to the 2 coherent receivers operating at 256~GSa/s. The 6 channels in each group are delayed in multiples of 30 m, corresponding to a 163-ns~delay. This setup allows for each coherent receiver to receive 6 simultaneously transmitted 200~GHz sub-channels each, enabling the demonstration of 12 sub-channels for a total of 2.4~THz bandwidth. Due to the TDM solution, the phase noise decorrelates slightly between consecutively received channels: we therefore we use a low linewidth ($<5$~kHz) laser as the comb pump laser. This is not required in a real system, provided the channels are sufficiently length-matched, so the comb can be pumped with a more typical (e.g.100~kHz) linewidth laser.

The captured samples are upsampled offline to an aggregate sampling rate of 3.072~TSa/s and aligned by cross correlating the overlapping spectral region (12.5 GHz) of each sub-channel, as indicated by the vertical red dashed lines in Fig.~\ref{fig:results}(b), which plots the 2.4~THz bandwidth spectrum observed by the receiver. The time offset between each channel is relatively stable between captures, but can experience some drift in this experiment due to the long lengths of fiber required for the TDM technique. Each of the 80-Gbd signals is then processed via standard coherent digital signal processing (DSP), and a 4$\times$2 multiple-input-multiple-output (MIMO) 801-tap least-means-squares (LMS) equaliser embedded with a digital phase locked loop. The MIMO equaliser compensates residual phase delay between both neighbouring channels and the I/Q components of the received signals~\cite{shi2017246}.
% Note this is emulating the processing in a optically parallelled THz receiver where the analog electronics are opearting at lower speed but the digital signal can be proceeed at the full speed of the receiver.
\section{Results and Discussion}
\begin{figure*}[t]
   \centering
    \input{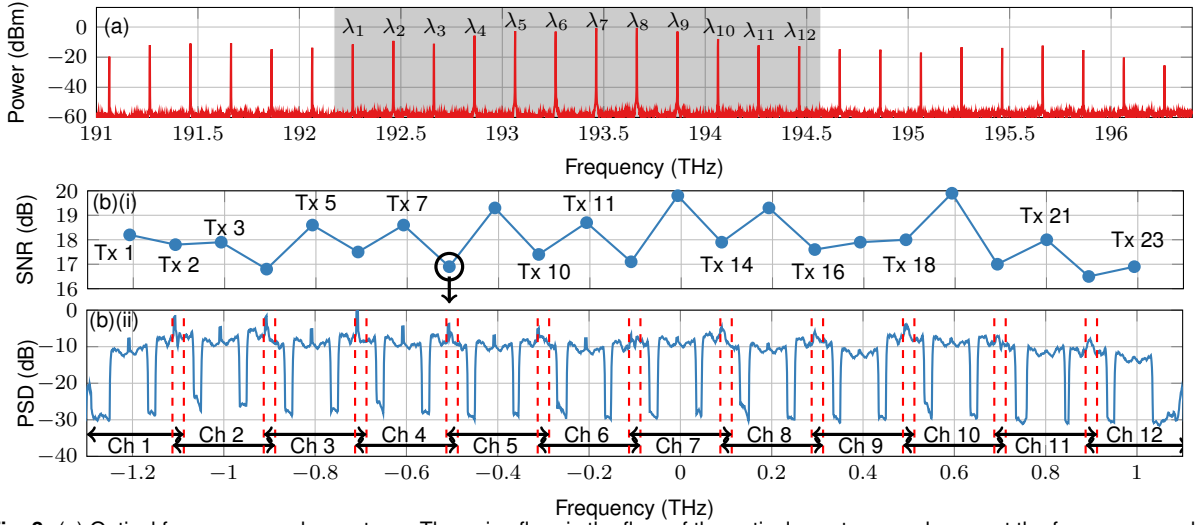}
    {\vspace{-0.2cm}}
    \caption{(a) Optical frequency comb spectrum. The noise floor is the floor of the optical spectrum analyser, not the frequency comb. (b)(i) Received SNR and (b)(ii) 2.4 THz coherent receiver spectrum after stitching together the 12 parallel sub-channels (Ch $n$). Red dashed lines indicate channel edges.}
    \label{fig:results}

\end{figure*}
\begin{figure}[htb]
   \centering
    \input{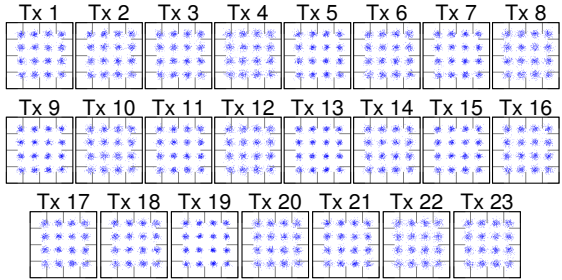}
    {\vspace{-0.2cm}}
    \caption{Constellations for the received test channels (Tx $m$).}
    \label{fig:constellations}

\end{figure}
Fig.~\ref{fig:results}(b)(ii) shows the recovered 2.4~THz spectrum after restitching, with recovered SNRs from the observed 23 Tx channels plotted above in Fig.~\ref{fig:results}(b)(i). The corresponding constellations are shown in Fig.~\ref{fig:constellations}. The SNR varies between 16.5~dB (Tx 22) and 19.9~dB (Tx 19), with a mean SNR of 18.1~dB. Part of the SNR variation is from the transmitter EDFA nonlinearities. This is because we modulate the 23 channels with a single IQ modulator, therefore the 23 ECLs are amplified via one EDFA before the modulator, resulting in a relatively strong four-wave mixing (FWM) in the EDFA. This is an artifact of the setup and would not exist if the channels were independently generated as in a normal telecom system. Despite the Tx nonlinearity, the SNRs of the sequences that require stitching (i.e. Tx 2, 4, \dots, 22) are still systematically around 1-3~dB lower than neighboring sequences that are captured fully within a signal sub-channel. This is for two reasons. Firstly, the poor roll-off and insertion loss of the WSS used in this experiment causes significant loss and distortion at the channel overlap, which could be improved by using a specially designed arrayed waveguide grating or other on-chip demultiplexing device. Secondly, the reduced responsivity of the photodiodes at the edge of the sub-channel reduces the receiver's sensitivity in these spectral regions, which is already limited by the power damage threshold of the photodiodes, since 6 channels must be sent to each coherent receiver due to the TDM technique used in this proof of concept experiment. 

Despite our non-ideal transmitters as well as the degradations brought by the TDM receiver, all the received 80-Gbd signals have an SNR of higher than 16.5 dB covering a total bandwidth of 2.4 THz, a significant improvement on the previous record of 610 GHz~\cite{drayss2023non} and the existing state of the art as plotted in Fig.\ref{fig:comparison}. The worst case pre-FEC BER was observed to be $1.3\times10^{-3}$ which is sufficient to allow error free ($<10^{-15}$) transmission for several hard-decision FEC schemes with overheads of 6.7\%~\cite{agrell2018information}. Therefore, despite zero effort to optimise the modulation format or coding for the channel, this represents a net bit rate of 6.87~Tb/s on a single polarisation, far beyond the previous record for a single polarisation coherent receiver (1.79~Tb/s~\cite{drayss2023non}).This highlights the huge potential microcombs have in scaling the capacity of a single optical receiver, enabling energy and cost efficient networks based on optical superchannels.% and the real-time, continuous analysis of full field optical waveforms at the femtosecond level.

\section{Conclusion}
We demonstrate a record 2.4-THz bandwidth coherent receiver using a photonic crystal microring resonator in tantalum pentoxide. These PhCRs offer stable soliton formation and high efficiencies, paving the way for efficient coherent reception of THz of bandwidth in a single optical transceiver.% and real-time, continuous analysis of full field optical waveforms at the femtosecond level.

%In addition to applications in telecommunications, real-time, continuous analysis of full field optical waveforms at the femtosecond level may allow for new insights into transient optical phenomena, with applications in solid state quantum information processing and quantum chemistry[CITE].

%Furthermore, the scheme demonstrated here can be converted into a photonic ADC by simply adding an Mach-Zehnder modulator with sufficient bandwidth~\cite{fang2021320}. These photonic ADCs could achieve THz(?) bandwidth using plasmonic and thin film lithium niobate modulators, enabling ADCs that vastly outperform their electronic counterparts on an effective jitter and energy per bit basis~\cite{deakin2023energy}.

%-------------------------------------------------- Acknowledgements Section -------------------------------------------------------%
\clearpage
%\section{Acknowledgements}
%One extra page is allowed so that acknowledgements and references can be given in full length.

%-------------------------------------------------- Bibliography Section -------------------------------------------------------%
% see also https://tex.stackexchange.com/questions/55030/text-before-references-but-after-bibliography-title-with-bibtex as of 2024-02-29
%\defbibnote{myprenote}{%
%Citations must be easy and quick to find. More precisely:
%\begin{itemize}
%    \item Please list all the authors. 
%    \item The title must be given in full length. 
%    \item Journal and conference names should not be abbreviated but rather given in full length.
%    \item The DOI number should be added incl. a link.
%\end{itemize}
%}
%\printbibliography[prenote=myprenote
\printbibliography
\vspace{-4mm}

%\newpage

%\begin{figure*}[t]
%   \centering
%    \input{figs/results_fig2}
%    {\vspace{-0.2cm}}
%    \caption{xxx}
%    \label{fig:results}

%\end{figure*}

%%%%%%%%%%%%%%%%%%%%%%%%%%%%%%%%%%%%%%%%%%%%%
%---------------------------------------------- End of Document -----------------------------------------------%
\end{document}